\documentclass[runningheads]{llncs}

\usepackage{eccv}

\usepackage{eccvabbrv}

\usepackage{graphicx}
\usepackage{booktabs}
\usepackage{xcolor}

\usepackage[accsupp]{axessibility}  

\newcommand{\comment}[1]{}

\usepackage{hyperref}

\usepackage{orcidlink}
\usepackage{array,multirow,graphicx}

\begin{document}

\newcolumntype{C}[1]{>{\centering\arraybackslash}p{#1}}

\title{Art2Mus: Bridging Visual Arts and Music through Cross-Modal Generation} 

\titlerunning{Art2Mus}

\author{Ivan Rinaldi\orcidlink{0009-0003-9597-064X} \and
Nicola Fanelli\orcidlink{0009-0007-6602-7504} \and
Giovanna Castellano\orcidlink{0000-0002-6489-8628} \and
Gennaro Vessio\orcidlink{0000-0002-0883-2691}}

\authorrunning{I.~Rinaldi et al.}

\institute{Department of Computer Science, University of Bari Aldo Moro, Italy
\email{name.surname@uniba.it}}

\maketitle

\newcommand{\ArtToMus}{\mathcal{A}\textit{rt2}\mathcal{M}\textit{us}}
\newcommand{\ArtToMusTok}{\mathcal{A}\textit{rt2}\mathcal{M}\textit{us}\text{-}\textit{4}}

\newcommand{\ArtToMusEDescTok}{\mathcal{A}\textit{rt2}\mathcal{M}\textit{us}\text{-}\textit{E}\text{\_}\textit{D}\text{-}\textit{1}\mathcal{T}\textit{ok}}
\newcommand{\ArtToMusEDescToks}{\mathcal{A}\textit{rt2}\mathcal{M}\textit{us}\text{-}\textit{E}\text{\_}\textit{D}\text{-}\textit{4}\mathcal{T}\textit{ok}}

\newcommand{\ArtGraph}{\mathcal{A}\textit{rt}\mathcal{G}\textit{raph}}
\newcommand{\FMA}{\mathcal{F\textit{ree}\ M\textit{usic}\ A\textit{rchive}}}

\newcommand{\nicola}[1]{\textcolor{orange}{Nicola: #1}}
\newcommand{\gio}[1]{\textcolor{magenta}{Prof: #1}}
\newcommand{\rino}[1]{\textcolor{blue}{Rino: #1}}
\newcommand{\ivan}[1]{\textcolor{red}{Ivan: #1}}

\begin{abstract}
Artificial Intelligence and generative models have revolutionized music creation, with many models leveraging textual or visual prompts for guidance. However, existing image-to-music models are limited to simple images, lacking the capability to generate music from complex digitized artworks. To address this gap, we introduce $\ArtToMus$, a novel model designed to create music from digitized artworks or text inputs. $\ArtToMus$ extends the AudioLDM~2 architecture, a text-to-audio model, and employs our newly curated datasets, created via ImageBind, which pair digitized artworks with music. Experimental results demonstrate that $\ArtToMus$ can generate music that resonates with the input stimuli. These findings suggest promising applications in multimedia art, interactive installations, and AI-driven creative tools. The code is publicly available at: \url{https://github.com/justivanr/art2mus_}.\footnote{Presented at the AI for Visual Arts (AI4VA) workshop at ECCV 2024.}

\keywords{Generative AI \and Image-to-Music Synthesis \and Music Generation \and Visual Arts}
\end{abstract}

\section{Introduction}

Generating music based on artworks represents a complex and largely unexplored challenge. Existing music generation models have demonstrated the ability to create music from textual prompts, additional audio samples, or simple real-world images~\cite{audioldm2,m2ugen}. However, none of the current state-of-the-art models use artworks in the music generation process. The ability to generate music from artworks could revolutionize multimedia experiences, offering new avenues for creative expression and interactive art.

The complexity of artwork-based music generation is exacerbated by the intricate nature of art, which requires models to translate visual artistic elements into musical components effectively. Factors such as color, subject matter, and style significantly influence the music to be generated, and capturing these nuances presents a significant challenge. Despite advances in deep learning methods for image information extraction, using these techniques to guide music generation from artworks remains an unexplored area.

This paper addresses these challenges by introducing $\ArtToMus$, a novel deep learning model designed to generate music from digitized artworks, their textual descriptions,  or a combination of both. Our model aims to bridge the gap in current research by leveraging the rich, detailed information inherent in artworks to produce corresponding musical compositions. To overcome the lack of existing digitized artwork-music pairs, we create two new synthetic datasets using ImageBind~\cite{imagebind} and LLaVA~\cite{llava}. $\ArtToMus$ leverages these datasets to produce music that resonates with the input artworks, pushing the boundaries of how humans create and experience music.

The rest of this paper is structured as follows. Section~\ref{sec:related} reviews related work in music generation. Section~\ref{sec:datasets} describes the datasets' creation process. Section~\ref{sec:art2mus} details the architecture and methodology of $\ArtToMus$. Section~\ref{sec:experiments} presents experimental results and analysis. Finally, Section~\ref{sec:conclusion} discusses implications of our findings and suggests future research directions.

\section{Related Work}
\label{sec:related}

Conditioned music generation is an emerging field that involves creating music based on specific input conditions, including text, audio, and images. Several models have been developed to tackle this challenge, leveraging deep learning techniques.

MusicLM~\cite{musiclm} generates music from text input using a hierarchical sequence-to-sequence architecture with Transformers~\cite{vaswani2017}. It is trained on a dataset of music-text pairs and can output music at 24 kHz. MusicLM utilizes several models: MuLan~\cite{mulan2022} to align text and music in an embedding space, SoundStream~\cite{soundstream2022} for high-quality audio compression, and w2v-BERT~\cite{w2vbert2021} for self-supervised speech representation learning.

MusicGen~\cite{musicgen} generates music based on either text, a melody, or both. Textual descriptions guide the style and mood, while melodic features ensure specific harmonic and melodic structures are matched. It employs a single-stage Transformer decoder architecture that works on compressed discrete music representation, producing 32 kHz music ranging from short snippets to several minutes.

AudioLDM~2~\cite{audioldm2} is a general-purpose audio generation model that includes speech, music, and sound effects. It introduces a general representation of audio called the ``language of audio'' (LOA), which consists of a sequence of vectors representing the semantic information of an audio clip. The input information is translated into the LOA representation for the model to generate audio, which can be conditioned by text or audio.

M\textsuperscript{2}UGen~\cite{m2ugen} integrates large language models with multi-modal encoders to generate music based on text, images, and videos. The input data is processed with the encoders, whose outputs are aligned and combined before being fed to Llama~2~\cite{llama2} to generate a text prompt reflecting the user's intent. The prompt is then fed to a music model to generate the desired audio.

Vis2Mus~\cite{vis2mus} is a system that explores the relationship between visual features and musical elements through visual-to-music mapping. It applies image transformations, such as adjusting brightness, contrast, or style transfer, to an input image and its corresponding piece of music. These transformations ensure that changes in the image's visual characteristics are reflected in the musical output, allowing users to modify existing music pieces based on visual alterations.

All the models mentioned above primarily generate music from text prompts, with some using additional audio tracks for conditioning. Among these, only M\textsuperscript{2}UGen accepts images as conditioning information, but it is trained on general real-world images, not including artworks. Vis2Mus also accepts images as input but allows users to modify existing music pieces rather than generate new ones. To our knowledge, the proposed $\ArtToMus$ is the first attempt to generate music using digitized artworks for conditioning.

\section{Proposed Datasets}
\label{sec:datasets}

This section provides an overview of the creation of the two datasets used to train, validate, and test our $\ArtToMus$ system. 

\subsection{Data Sources}

In the first phase, we sought unrestricted artwork and music datasets. We aimed to create datasets of paired digitized artworks and music tracks, which would serve as our ground truth in the experimental setting.

$\ArtGraph$~\cite{castellano2022}, an artistic knowledge graph based on WikiArt and DBpedia, served as our artwork data source. It includes 116,475 artworks across 18 genres and 32 styles, making it ideal for our needs. In addition to digitized images, it provides extensive metadata such as genre, style, and artist. Due to data volume, we worked with a subset of 10,000 digitized artworks from $\ArtGraph$, preserving all styles and genres.

Free Music Archive (FMA)~\cite{defferrard2016}, a dataset designed for music information retrieval tasks, served as our music data source. It contains music tracks and metadata from a free and open library dump directed by WFMU, the longest-running freeform radio station in the United States. We utilized the Large FMA dataset for our study, which includes around ~100 GB of music data, all licensed under Creative Commons, consisting of 106,574 music files, each 30 seconds long, from 16,341 artists, categorized under a hierarchy that includes 16 top genres and 161 genres. Some music tracks were corrupted, causing issues with ImageBind. Consequently, 689 music files were excluded from the subsequent processing phases, leaving 105,885 from the initial 106,574 files.

\subsection{Dataset Creation}

To pair digitized artworks and music tracks, we employed ImageBind~\cite{imagebind}. This model uses image-paired data to learn joint embeddings across six different modalities (images, text, audio, depth, thermal, and IMU data). After preprocessing, we fed the data into ImageBind for inference, obtaining a single tensor embedding for each input type. Using cosine similarity, we effectively compared and associated music with digitized artworks.

Specifically, we created two distinct datasets:

\begin{enumerate} 
    \item \textit{Artwork-music dataset}: This dataset includes 10,000 unique digitized artworks and 10,000 unique music tracks. 
    Each artwork was paired with the most similar music track based on the cosine similarity between their embeddings. After pairing, the selected music track was excluded to avoid multiple assignments.
    \item \textit{Artwork-description dataset}: For each artwork in the \textit{artwork-music dataset}, we generated a description using LLaVA~\cite{llava}, a multimodal large language model. LLaVA was fed with the artwork's digitized image and the simple textual prompt ``\textit{Describe this artwork}'' to generate the artwork description. Subsequently, music tracks were assigned to these descriptions based on the artworks they described. This resulted in a dataset containing 10,000 pairs of artwork descriptions and music tracks.
\end{enumerate}

The \textit{artwork-music dataset} has been designed to evaluate $\ArtToMus$'s ability to generate music directly from artworks. In contrast, the \textit{artwork-description dataset} has been proposed to assess the performance of AudioLDM~2 when generating music using text descriptions alone. 


\section{$\ArtToMus$}
\label{sec:art2mus}

\begin{figure}[t]
  \centering
    \includegraphics[width=\linewidth]{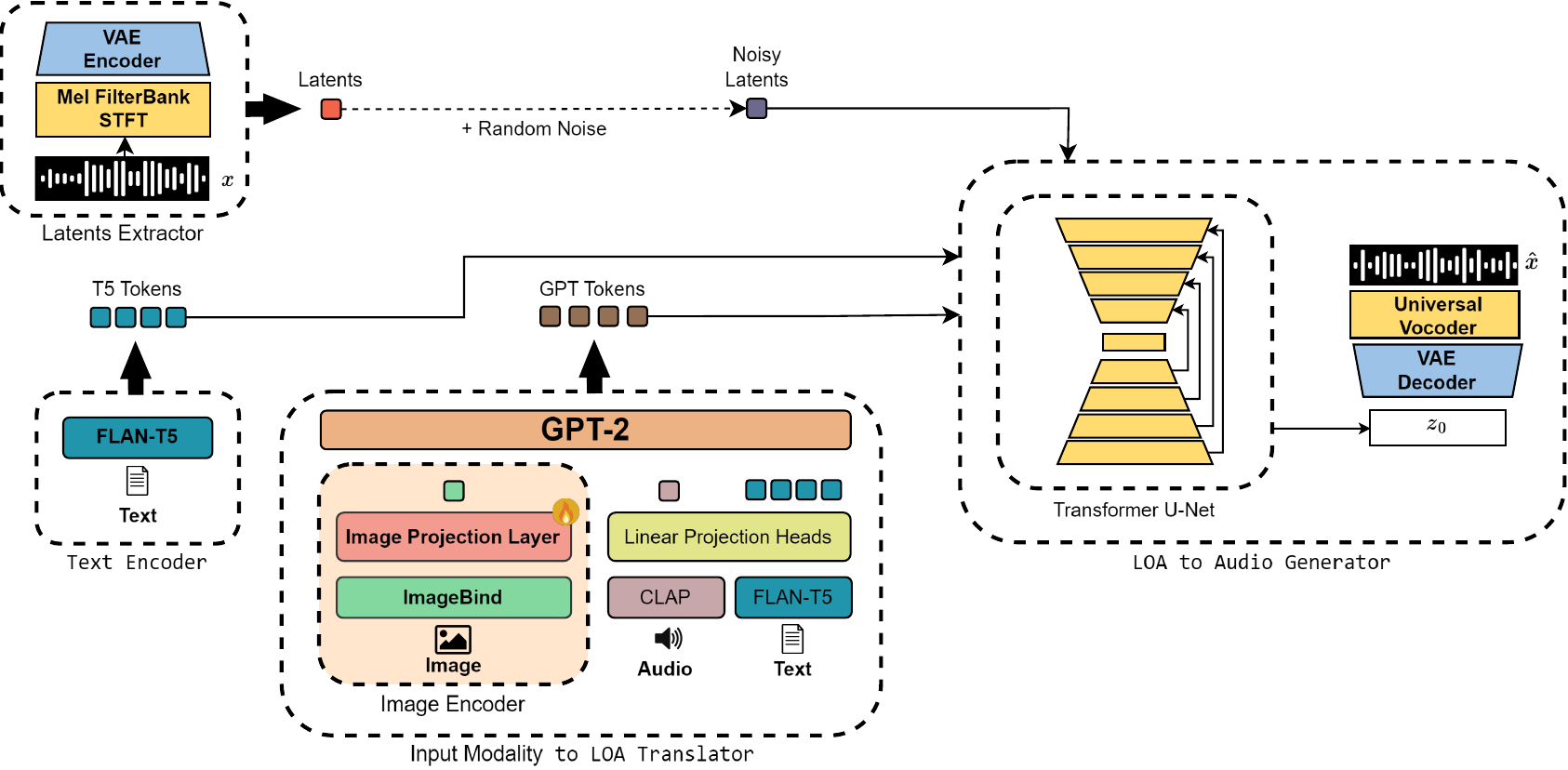}
    \caption{An overview of $\ArtToMus$. We extend the AudioLDM~2 model by incorporating an \textit{Image Encoder}, which consists of the ImageBind vision model and a trainable \textit{Image Projection Layer}. The rest of the architecture is initialized with an AudioLDM~2 checkpoint and remains frozen. GPT-2 acts as a language of audio (LOA) translator, converting text and image embeddings into conditioning information for a latent diffusion model. Additional conditioning comes from a \textit{Text Encoder} using FLAN-T5. Music is converted into mel-spectrograms by the \textit{Latent Extractor}, and synthesized into audio by the \textit{LOA to Audio Generator}, integrating all inputs.}
    \label{fig:art2mus_architecture}
\end{figure}

In this section, we introduce $\ArtToMus$, a model capable of generating music from digitized artworks (or text). The proposed model leverages the AudioLDM~2 architecture, which supports text input for conditioning (Fig.~\ref{fig:art2mus_architecture}). This work extends the AudioLDM~2 architecture to allow digitized artworks to guide music generation.

The AudioLDM2 architecture includes the following key components: AudioMAE~\cite{audiomae} for extracting language of audio (LOA) representations (introduced in~\cite{audioldm2}), GPT-2~\cite{gpt2} for translating input conditions into LOA, and a latent diffusion model (LDM)~\cite{latent_diff_models} for generating audio waveforms. LDM is used to denoise VAE-encoded mel-spectrograms over multiple timesteps, utilizing LOA and text embeddings as conditioning inputs. This process generates denoised latent representations, further decoded into spectrograms and subsequently into new synthetic audio. GPT-2 and LDM are pre-trained on unlabeled audio data to learn general audio features. Despite its effectiveness in audio generation tasks such as music, speech, and sound effects when conditioned on textual information, AudioLDM 2 cannot directly use images as input. Consequently, it cannot use digitized images of artworks to guide the audio generation process.

\begin{figure}[t]
  \centering
    \includegraphics[width=0.6\linewidth]{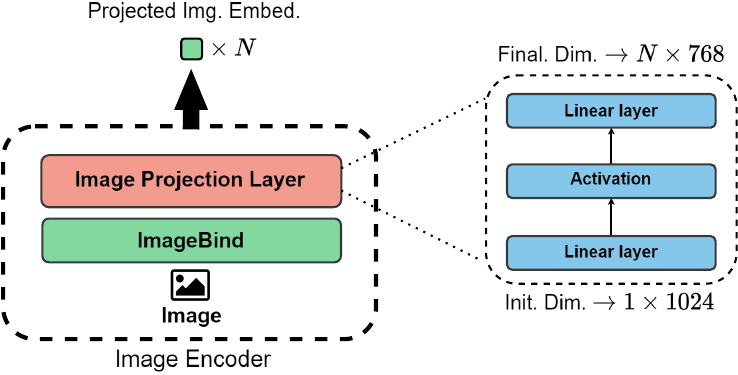}
    \caption{The \textit{Image Encoder}. ImageBind processes an artwork image to generate an image embedding. This embedding is then linearly transformed by the \textit{Image Projection Layer}, which consists of two linear layers and an activation layer to match the GPT-2 embedding dimension. The encoder outputs $N$ tokens, each with a dimension of 768. \label{fig:image_encoder}}
\end{figure}

To handle artwork images, in addition to text, we extended AudioLDM~2 by introducing an additional component called \textit{Image Encoder}. This component generates the image embedding of a digitized artwork using ImageBind. The embedding is then processed by the \textit{Image Projection Layer} to align with the embedding dimensions required by the GPT-2 model, producing a single token fed into GPT-2. Figure~\ref{fig:image_encoder} illustrates the \textit{Image Encoder}, which is located within the \textit{Input Modality to LOA Translator} just before the GPT-2 model. Additionally, we pass some text to the \textit{Text Encoder} for music generation from digitized artworks. It encodes a textual prompt using the FLAN-T5 encoder~\cite{flant5}. The encoded prompt is then fed to a T-UNet~\cite{audioldm2}, along with the tokens generated by GPT-2, providing additional guidance to the generation process.

Music paired with the digitized artwork is processed by the \textit{Latent Extractor} using a Short-Time Fourier Transform module~\cite{tacotronstft} to convert the music track into a mel-spectrogram. This mel-spectrogram is then encoded using the VAE from AudioLDM~2's architecture. Latent features are extracted from the VAE's output, and are then combined with randomly picked noise, resulting in noisy latents. Finally, the \textit{LOA to Audio Generator} employs the LDM from AudioLDM~2's architecture along with the combined output from these components—the GPT-2 tokens, the FLAN-T5 tokens, and the noisy latents—to generate music tailored to the artwork. LDM uses a T-UNet to predict the noise in the noisy latents, which is then removed to achieve clean latents. Using the VAE's decoder, the T-UNet's output is decoded into a spectrogram, and then converted into a waveform, corresponding to the generated music.

During training, $\ArtToMus$ learns from triplets of data made of three elements: the conditioning information (digitized artwork, text, or both), a music track, and a predefined sentence guiding the generation process to achieve the desired result (``Music representing the content of this artwork''). The \textit{Image Encoder} uses ImageBind to embed input digitized artworks, followed by a linear transformation through the \textit{Image Projection Layer}. The processed embedding is then fed to GPT-2 to generate tokens. Concurrently, the \textit{Text Encoder}, employing FLAN-T5, generates text embeddings from the sentence ``Music representing the content of this artwork''.

Along with the triplets of data, $\ArtToMus$ requires a negative prompt to guide the generation process and avoid undesired outputs (e.g., a negative prompt such as ``Metal music'' guides $\ArtToMus$ to generate music distinct from metal music).

\section{Experiments}
\label{sec:experiments}

This section details the training, validation, and conducted experiments for $\ArtToMus$. We provide the specific hyperparameter setting and discuss the performance metrics. Finally, we present the experimental results.

\subsection{Exploratory Analysis}

Before training, we analyzed the synthetic datasets to assess the similarity between elements in each pair. This assessment was conducted by calculating the cosine similarity between the ImageBind embeddings of the paired elements. Table~\ref{tab:pairwise_similarities_summary} summarizes the results, including statistics such as each dataset's average, maximum, and minimum pairwise similarity values.

The highest similarity value achieved is $0.4692$ in the \textit{artwork-music dataset}, while the lowest is $-0.0954$ in the \textit{artwork-description dataset}. The \textit{artwork-music dataset} also reports the highest average value at $0.2746$, compared to the \textit{artwork-description dataset}'s average of $0.0501$. Additionally, both datasets contain more pairs with similarities below the average, with the \textit{artwork-music dataset} showing the most significant disparity (728 more pairs below the average). These findings highlight different degrees of similarity within the created synthetic datasets, which could influence the training process and the model's performance. As for the provided statistics, the different similarity values suggest that further refinement in the dataset creation phase is needed to ensure higher-quality training sets.

\begin{table}[t]
  \centering
  \caption{Summary of pairwise similarity scores. The table includes the maximum, minimum, and average similarity values for each dataset, as well as the distribution of pairs with scores above and below the average similarity.}
  \begin{tabular}{lccccc}
    \hline
    \textbf{Dataset} & \textbf{Max sim.} & \textbf{Min sim.} & \textbf{Avg.~sim.} & \textbf{Above avg.} & \textbf{Below avg.} \\
    \hline
    Artwork-music & 0.4692 & 0.1532 & 0.2746 & 4636 & 5364 \\
    Artwork-description & 0.3569 & --0.0954 & 0.0501 & 4942 & 5058 \\
    \hline
  \end{tabular}
  \label{tab:pairwise_similarities_summary}
\end{table}

\subsection{Training \& Validation}

$\ArtToMus$ was trained on a single NVIDIA RTX 4090 24GB GPU, requiring approximately 8 to 21 GB of VRAM. 
During training, only the \textit{Image Projection Layer} in the \textit{Image Encoder} was kept unfrozen to allow its weights to be updated. We used the AdamW optimizer with default hyperparameters, a learning rate of 2e--5, and a constant learning scheduler with 500 warmups and 10,000 learning steps. $\ArtToMus$ was trained for 20 epochs with a batch size of 4 and 4 gradient accumulation steps. The \textit{artwork-music dataset} was used for all training, validation, and testing phases, divided into an 80/20 split for training and testing, with 100 instances extracted from the test set for validation due to the time-consuming validation process. The artworks' styles stratified the splitting. During validation, a single music track of 10 seconds was generated with 200 inference steps, using ``Low quality'' as the negative prompt.

We computed the Mean Squared Error (MSE) loss and the Signal to Noise Ratio (SNR) loss during training. MSE measures the average squared differences between predicted and actual values, penalizing larger errors more. It ensures non-negative values, with a lower MSE indicating better model performance. It is calculated as: 
\begin{equation}
\text{MSE} = \frac{1}{n} \sum_{i=1}^{n} (y_i - \hat{y}_i)^2
\label{eq:mse}
\end{equation}
where \( y_i \) represents the noisy latents, consisting of \( n \) features extracted from the ground truth music for the \( i \)-th artwork, while \( \hat{y}_i \) represents the noise, consisting of \( n \) features of the music generated by the model for the \( i \)-th artwork.

SNR~\cite{snr_gamma} measures the desired signal level against background noise. Our approach calculates the SNR for each timestep using $\ArtToMus$'s scheduler. This SNR is then used to weigh the MSE loss: higher SNR values (clearer audio) receive more weight, while lower SNR values (noisier audio) receive less weight. This ensures the model focuses more on learning from more precise data. SNR is calculated as:
\begin{equation}
\text{SNR} = \frac{\alpha^2}{\sigma^2}
\label{eq:snr}
\end{equation}
where \( \alpha \) is the remaining signal from the original music piece at a specific timestep, while \( \sigma \) is the noise added to the music during the generation process. A higher \( \alpha \) means more original music is preserved, while a higher \( \sigma \) means more noise is present. Squaring the ratio gives the power ratio of the signal to the noise, the standard way to represent SNR, reflecting how much stronger the signal is compared to the noise. The computed SNR value is combined with an SNR gamma (5.0) to adjust the MSE loss weights. The final loss is a weighted average of the MSE loss.

\subsection{Performance Metrics}

We computed the Kullback-Leibler divergence (KL-Div), the Frechet Audio Distance (FAD) score~\cite{kilgour2018frechet}, and the ImageBind score (IBSc) during evaluation. 

KL-Div measures the dissimilarity between two probability distributions $\mathcal{P}$ and $\mathcal{Q}$. A lower KL-Div means the predicted distribution is closer to the true distribution. It is calculated as: 
\begin{equation}
\text{KL-Div}(\mathcal{P} \| \mathcal{Q}) = \sum_{i} \mathcal{P}(i) \log \left( \frac{\mathcal{P}(i)}{\mathcal{Q}(i)} \right)
\label{eq:kl_div}
\end{equation}
where $\mathcal{P}$ is the ground truth music distribution, and $\mathcal{Q}$ is the generated music distribution for the \( i \)-th artwork.

FAD measures the similarity between an audio clip and clean, studio-recorded music, using VGGish (a pre-trained audio classification model~\cite{vggish}) embeddings. A higher FAD score indicates greater dissimilarity. It is calculated as: 
\begin{equation}
\text{FAD} = \| \mu_b - \mu_e \|^2 + \text{Tr}\left( \Sigma_b + \Sigma_e - 2\sqrt{\Sigma_b \Sigma_e} \right)
\label{eq:fad_score}
\end{equation}
where $\mu_b$ and $\mu_e$ are the means of the two multivariate normal distributions computed from the embeddings of the considered music tracks, $\Sigma_b$ and $\Sigma_e$ are their covariance matrices, and $\text{Tr}$ is the trace of a matrix. 

IBSc is a custom metric we defined to assess the similarity between a digitized artwork or its associated ground truth music track and the generated music track. We compute their ImageBind embeddings and calculate the cosine similarity between them. The IBSc is computed as follows: 
\begin{equation}
\text{IBSc}({emb}_{agt}, {emb}_{gen}) = \frac{{emb}_{agt} \cdot {emb}_{gen}}{\|{emb}_{agt}\| \|{emb}_{gen}\|}
\label{eq:cosine_sim_imagebind_score}
\end{equation}
where ${emb}_{agt}$ refers to either the digitized artwork embedding or the ground truth music track embedding, while ${emb}_{gen}$ represents the generated music track embedding.

Additionally, we involved a group of 10 participants for a subjective evaluation. Each participant was presented with 20 artworks (one at a time) along with music generated by two different models for each artwork. Participants rated the audio quality and pertinence of the music to the artwork, and chose which music track aligned more with the artwork, by answering questions as shown in Table~\ref{tab:subj_eval_questions}.

\begin{table}[t]
  \centering
  \caption{Subjective evaluation questions used in our experiments.}
  \resizebox{\linewidth}{!}{
    \begin{tabular}{lcc}
      \hline
      \textbf{Question} & \textbf{Music aspect} & \textbf{Scale} \\
      \hline
      How well do you think the music sounds? & Audio quality & 1--5 \\
      How pertinent is the music in your opinion? & Pertinence to artwork & 1--5 \\
      Which music track aligns more with the artwork? & Fitting to artwork & 1 or 2 \\
      \hline
    \end{tabular}
  }
  \label{tab:subj_eval_questions}
\end{table}

\subsection{Results}

We evaluated $\ArtToMus$ in image-to-music generation against 
AudioLDM~2, and AudioLDM~2-Music (with AudioLDM~2 capable of generating music, human speech, or sound effects from text, and AudioLDM~2-Music trained only for music generation). We used the \textit{artwork-music dataset} with $\ArtToMus$-based models and the \textit{artwork-description dataset} with AudioLDM~2 models to generate music. AudioLDM~2 models have been used as-is, without any tuning. The comparisons aimed to determine if $\ArtToMus$ could outperform its baseline (AudioLDM~2) and the specialized music-generating version of the baseline (AudioLDM~2-Music). Additionally, we compared $\ArtToMus$ with a modified version called $\ArtToMusTok$, which updates the \textit{Image Encoder} to output four tokens instead of one.

Table~\ref{tab:exp_res_obj_metrics} shows the average objective metrics for each model. The $\ArtToMus$-based models only surpass the KL-Div metric's baselines. It is worth noting that, although $\ArtToMus$ and $\ArtToMusTok$ do not outperform the baselines on FAD and IBSc, they are based on complex images of digitized artworks rather than merely text. Nevertheless, the base $\ArtToMus$ achieves the lowest KL-Div metric result ($0.002285$), and $\ArtToMusTok$ achieves better results—compared to $\ArtToMus$—for both FAD and IBSc scores, likely due to the additional tokens generated by the \textit{Image Encoder}. This observation is promising, as the increase in the IBSc between the digitized artwork and the generated music suggests that if we want to work with only digitized artworks, minor changes in the \textit{Image Encoder} will likely lead to better results.

For subjective metrics, Table \ref{tab:exp_res_subj_metrics} presents the average scores for each model. In the first experiment, $\ArtToMus$'s generated music is not preferred over AudioLDM~2's. In the second experiment, although $\ArtToMus$ generates higher quality music based on participants' evaluations, AudioLDM~2's generated music is still preferred. Similarly, although $\ArtToMusTok$ improved objective metrics, AudioLDM~2's generated music remains preferred. 

\begin{figure}[t]
    \centering
    \begin{minipage}{0.3421\textwidth}
        \centering
        \includegraphics[width=\textwidth]{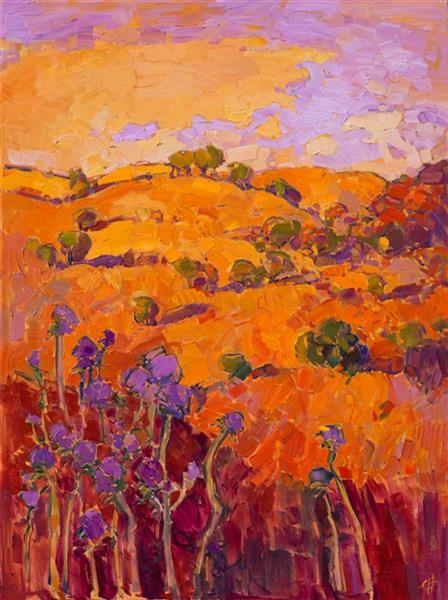}
        \vspace{0.001cm}
    \end{minipage}
    \hfill
    \begin{minipage}{0.64\textwidth}
        \centering
        \begin{subfigure}{\textwidth}
            \centering
            \includegraphics[width=\textwidth]{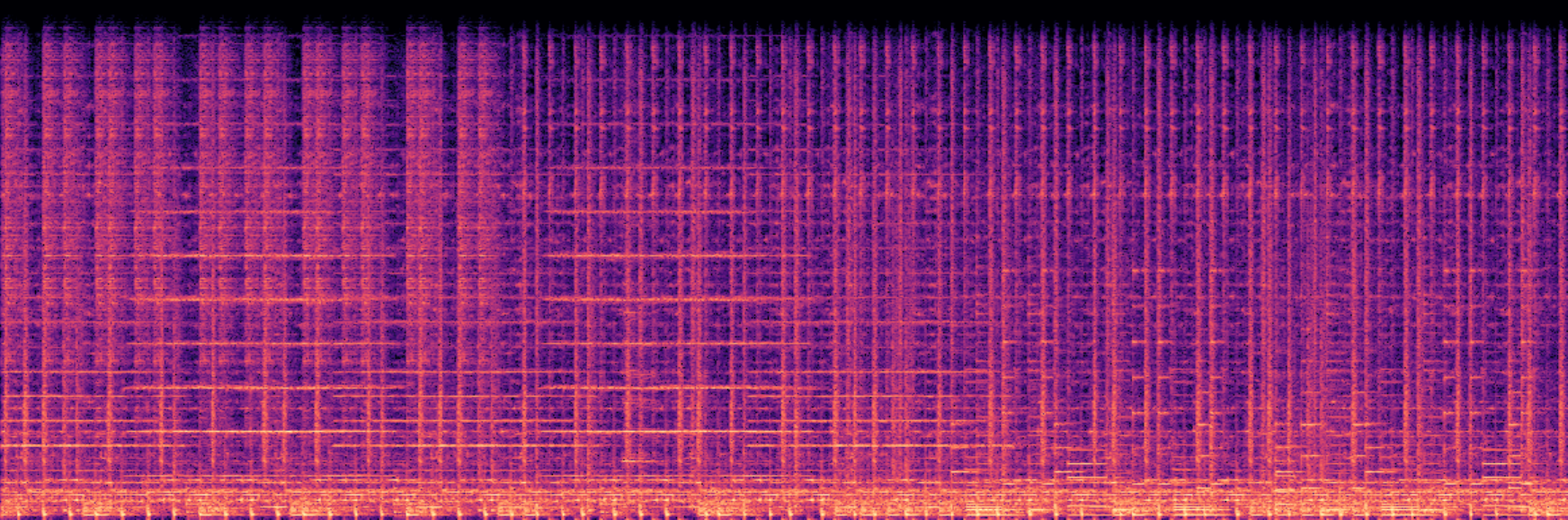}
            \caption{Ground truth music track's spectrogram}
            \label{fig:GT_SPECT_erin}
        \end{subfigure}
        \begin{subfigure}{\textwidth}
            \centering
            \includegraphics[width=\textwidth]{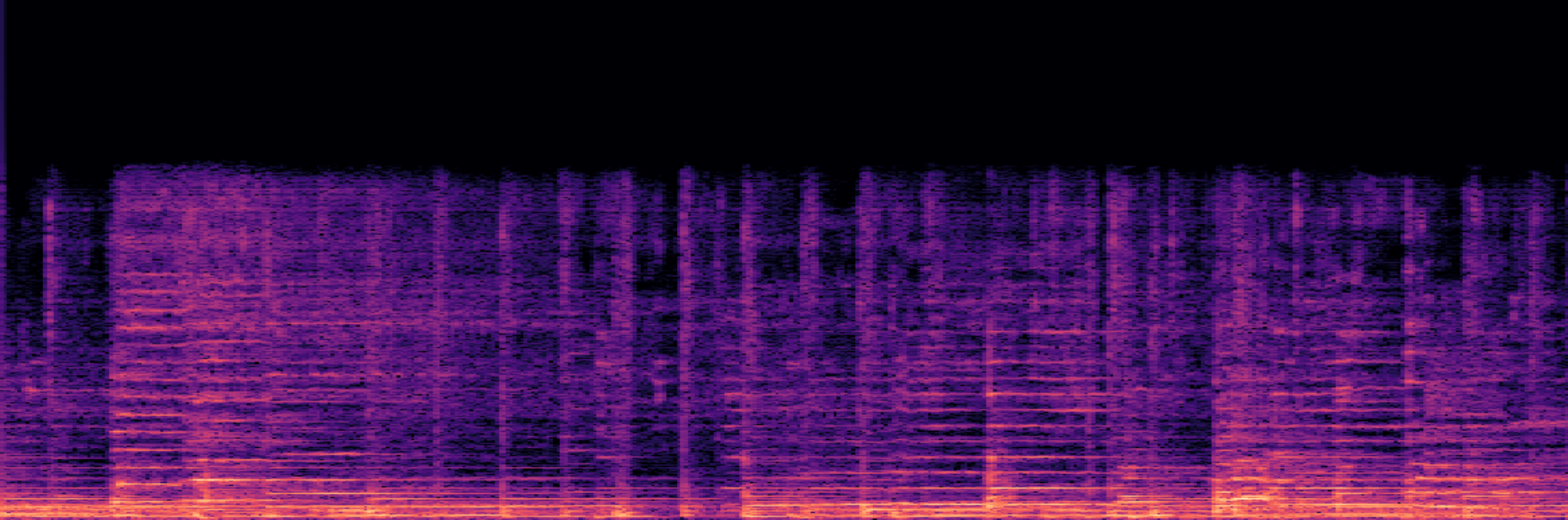}
            \caption{Generated music track's spectrogram}
            \label{fig:GM_SPECT_erin}
        \end{subfigure}
    \end{minipage}
    \caption{``Thistles on Orange'' by Erin Hanson, with associated spectrograms of the ground truth music track and the generated music track.}
    \label{fig:erin_hanson_thistles_on_orange_spectrograms}
\end{figure}

\begin{figure}[t]
    \centering
    \begin{minipage}{0.3467\textwidth}
        \centering
        \includegraphics[width=\textwidth]{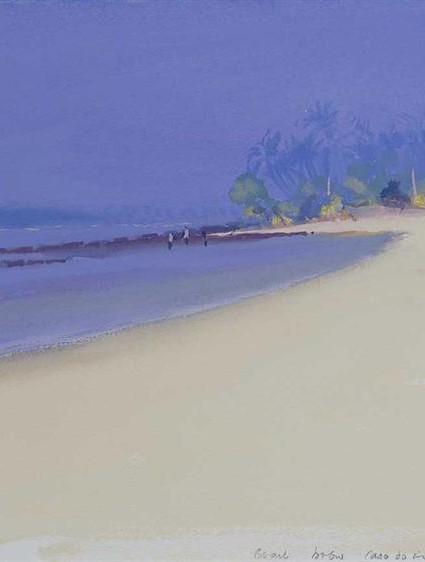}
        \vspace{0.00001cm}
    \end{minipage}
    \hfill
    \begin{minipage}{0.64\textwidth}
        \centering
        \begin{subfigure}{\textwidth}
            \centering
            \includegraphics[width=\textwidth]{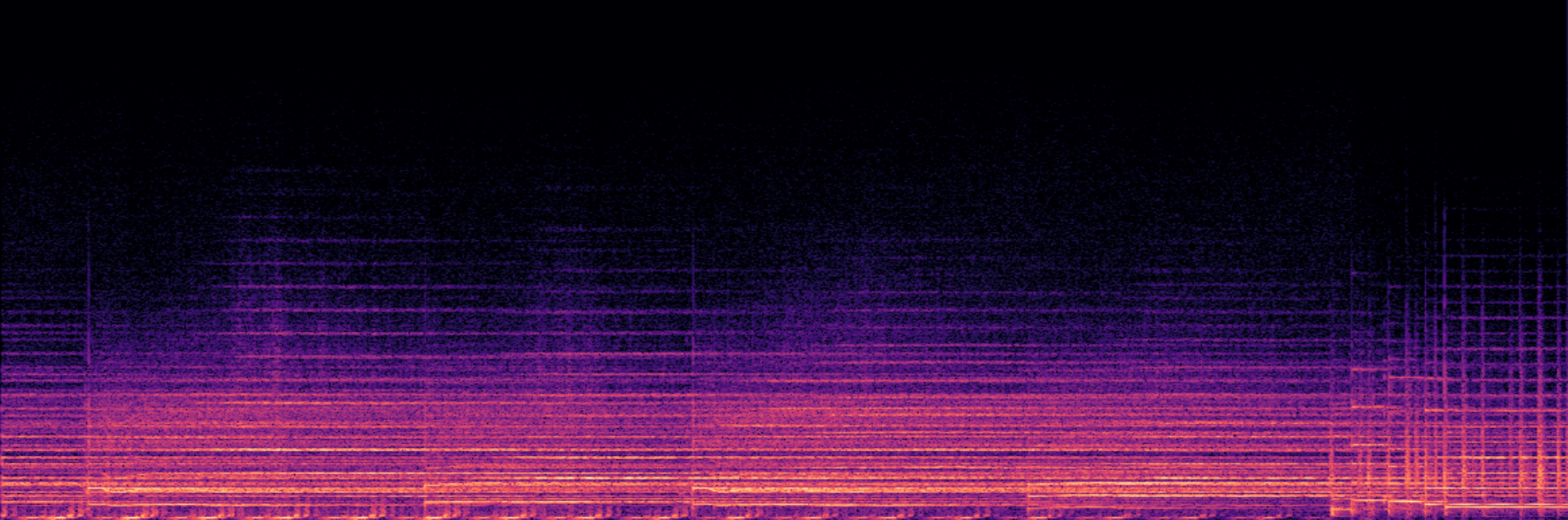}
            \caption{Ground truth music track's spectrogram}
            \label{fig:GT_SPECT_john}
        \end{subfigure}
        \begin{subfigure}{\textwidth}
            \centering
            \includegraphics[width=\textwidth]{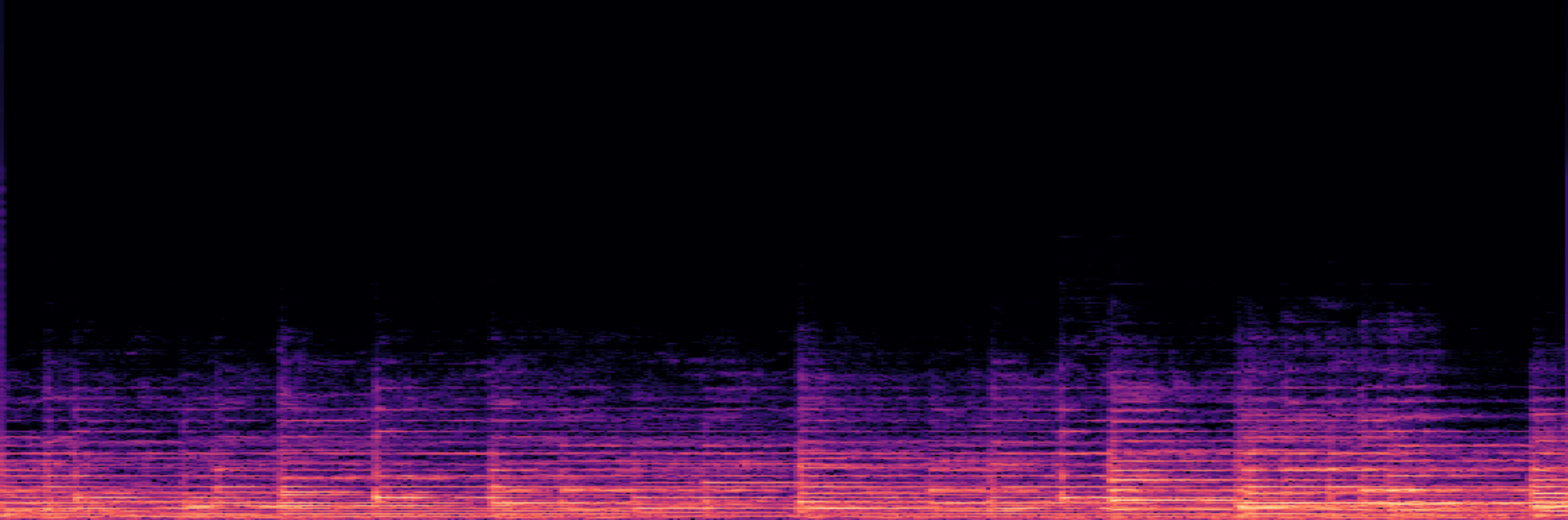}
            \caption{Generated music track's spectrogram}
            \label{fig:GM_SPECT_john}
        \end{subfigure}
    \end{minipage}
    \caption{``Casa do Rio Beach, Goa'' by John Miller, with associated spectrograms of the ground truth music track and the generated music track.}
    \label{fig:ohn-miller_casa_do_rio_beach_goa_spectrograms}
\end{figure}

Additionally, Figs.~\ref{fig:erin_hanson_thistles_on_orange_spectrograms}--\ref{fig:ohn-miller_casa_do_rio_beach_goa_spectrograms} compare the spectrograms of the ground truth music tracks associated with two different artworks and the music tracks generated based on these artworks using $\ArtToMus$. Examples of music generated with $\ArtToMus$ and the corresponding artworks are available on Google Drive.\footnote{\url{https://rb.gy/riwg2f}}

\begin{table}[t]
    \centering
    \caption{Summary of average objective metrics from experiments, including FAD, KL-Div, and IBSc. The arrows next to the metrics' names indicate whether a lower or higher value is desirable. Artw stands for artwork, GeMus stands for generated music, and GtMus stands for ground truth music.}
    \resizebox{\linewidth}{!}{
        \begin{tabular}{lcccc}
            \hline
            \textbf{Model} & \textbf{FAD $\downarrow$} & \textbf{KL-Div $\downarrow$} & \textbf{IBSc (Artw-GeMus) $\uparrow$} & \textbf{IBSc (GtMus-GeMus) $\uparrow$} \\
            \hline   
            $\ArtToMus$ & 22 & \textbf{0.002285} & --0.000474 & 0.100854 \\
            $\ArtToMusTok$ & 21 & 0.003082 & 0.007739 & 0.102437 \\
            AudioLDM~2 & \textbf{20} & 0.004107 & \textbf{0.097634} & \textbf{0.232370} \\
            AudioLDM~2-Music & 21 & 0.005168 & 0.083886 & 0.220064 \\
            \hline
        \end{tabular}
    }
    \label{tab:exp_res_obj_metrics}
\end{table}

\begin{table}[t]
    \centering
    \caption{Summary of average subjective metrics from experiments, including audio quality, pertinence to artwork and fitting to artwork (number of participants who preferred the model's generated music). The arrows next to the metrics' names indicate whether a lower or higher value is desirable.}
        \begin{tabular}{lccc}
            \hline
            \textbf{Model} & \textbf{Audio Quality $\uparrow$} & \multicolumn{2}{c}{\textbf{Compared to Artwork}} \\
            \cline{3-4}
            & & \multicolumn{1}{c|}{\textbf{Pertinence $\uparrow$}} & \textbf{Fitting $\uparrow$} \\
            \hline
            $\ArtToMus$ & 3.1 & 2.7 & 2/10 \\  
            AudioLDM~2 & \textbf{3.3} & \textbf{3.3} & \textbf{8/10} \\ 
            \hline
            $\ArtToMus$ & \textbf{3.1} & 2.7 & 4/10 \\
            AudioLDM~2-Music & 2.7 & \textbf{2.9} & \textbf{6/10} \\
            \hline
            $\ArtToMus$ & \textbf{3.1} & \textbf{2.7} & \textbf{7/10} \\
            $\ArtToMusTok$ & \textbf{3.1} & 2.5 & 3/10 \\ 
            \hline
        \end{tabular}
    \label{tab:exp_res_subj_metrics}
\end{table}

\subsection{Discussion}

Based on the experimental results, $\ArtToMus$-based models do not surpass the baselines. This may be attributed to the fact that  $\ArtToMus$ and $\ArtToMusTok$ are based only on pairs of complex images (digitized artworks) and music. This approach is more challenging than working with text, a more consolidated methodology~\cite{musiclm, musicgen, audioldm2, m2ugen}.

Among $\ArtToMus$-based models, the highest IBSc scores are achieved with $\ArtToMusTok$, with an IBSc of $0.007739$ between the digitized artwork and the generated music, and an IBSc of $0.102437$ between the ground truth music and the generated music. All models average an FAD of approximately 20, indicating variability in the audio quality. However, $\ArtToMus$-based models, especially $\ArtToMus$, achieve the lowest values for the KL-Div, suggesting that the generated music probability distribution aligns more closely with the ground truth than AudioLDM~2 models. 

Overall, the results indicate that $\ArtToMus$-based models generate music that does not align closely with the digitized artwork, while the remaining models (AudioLDM~2 and AudioLDM~2-Music) show better performance. 

Regarding subjective metrics, both $\ArtToMus$ and AudioLDM~2 models report scores that do not exceed 3.3 for the pertinence of the generated music to the artworks and the audio quality of the generated music. The results suggest that the generated music only slightly aligns with the artwork and could be better. Participant feedback indicates a preference for the music generated by AudioLDM~2 models over $\ArtToMus$. However, the latter experiment highlights the potential for architectural adjustments to improve $\ArtToMus$'s performance.

Apart from the base version of $\ArtToMus$, $\ArtToMusTok$ provides a foundation for future developments and improvements in music generation based solely on pairs of digitized artworks and music.

\section{Conclusion}
\label{sec:conclusion}

In this paper, we presented $\ArtToMus$, an innovative model for generating music based on digitized artworks, built on the AudioLDM~2 architecture. We introduce a key component, an \textit{Image Encoder}, which uses ImageBind to create embeddings from digitized artworks. Additionally, we utilize a \textit{Text Encoder}, which employs FLAN-T5 to generate text embeddings from designed sentences to guide the music generation process. Despite some limitations in producing high-quality music from digitized artworks, our results are promising and provide a strong foundation for future improvements.

We also introduced two new synthetic datasets, pairing digitized artworks from the $\ArtGraph$ knowledge graph with music tracks from the Free Music Archive dataset. While demonstrating potential, these datasets require further research to improve their size and quality, including incorporating more diverse artworks and music tracks.

Future work for $\ArtToMus$ includes exploring alternative embedding models and pairing techniques, utilizing larger and higher-quality datasets, and enhancing the \textit{Image Projection Layer} to capture the nuances of artwork details better. Additional metadata, such as historical, geographical, or emotional information, can enrich the datasets and lead to a more diverse and fitting music generation. Integrating models like LLaVA directly into the architecture could enable the generation of more detailed prompts beyond predefined ones, addressing aspects such as genre or rhythm and thereby improving the relevance and quality of the generated music. Finally, developing new and meaningful metrics to evaluate the generated music is crucial. 

Our approach opens new possibilities at the intersection of visual arts and music, offering innovative tools for artists and enriching the multimedia experience for audiences.

\bibliographystyle{splncs04}
\bibliography{main}

\end{document}